\documentclass[twocolumn,showpacs,prb]{revtex4}
\usepackage{graphicx}
\usepackage{dcolumn}
\usepackage{amsmath}
\usepackage{latexsym}
\usepackage{longtable}
\begin{document}
\title{Effect of size reduction on the ferromagnetism  of the manganite $\mathrm{La_{1-x}Ca_{x}MnO_3}$ (x=0.33)}
    
\author{Tapati Sarkar\footnote[1]{email:tapatis@bose.res.in} and  A. K. Raychaudhuri\footnote[2]{email:arup@bose res.in}}
\affiliation{DST Unit for NanoSciences, S.N.Bose National Centre for Basic Sciences, Block JD, Sector III, Salt Lake, Kolkata 700 098, West Bengal, India.}
\author{A. K. Bera and S. M. Yusuf}
\affiliation{Solid State Physics Division, Bhabha Atomic Research Centre, Mumbai 400 085, India.}

\date{\today}
\begin{abstract}

\noindent

In this paper we report an investigation on the ferromagnetic state and the nature of ferromagnetic transition of nanoparticles of $\mathrm{La_{0.67}Ca_{0.33}MnO_3}$ using magnetic measurements and neutron diffraction. The investigation was made on nanoparticles with crystal size down to $15$ nm. The neutron data show that even down to a size of $15$ nm the nanoparticles show finite spontaneous magnetization ($M_S$) although the value is much reduced compared to the bulk sample. We observed a non-monotonic variation of the ferromagnetic to paramagnetic transition temperature $T_C$ with size $d$ and found that $T_C$ initially enhances on size reduction, but for $d < 50$ nm it decreases again.  The initial enhancement in $T_C$ was related to an increase in the bandwidth that occured due to a compaction of the Mn-O bond length and a straightening of the Mn-O-Mn bond angle, as determined form the neutron data. The size reduction also changes the nature of the ferromagnetic to paramagnetic transition from first order to second order with critical exponents approaching mean field values. This was explained as arising from a truncation of the coherence length by the finite sample size.
\end{abstract}
\maketitle

\noindent
\section{\bf INTRODUCTION} 
The effect of size reduction on physical properties of hole doped manganites is a topic of considerable interest. The properties of nanoparticles of manganites (often  called "nanomanganites")  have been investigated recently.\cite{TKNath,NM2,NM3,NM4,NM5,NM6,NM7,NM8,NM9,Lopez} In general, properties of nanoparticles of magnetic materials can arise from intrinsic properties, or from factors arising from interactions of nanoparticles.\cite{MN1,MN2} In this paper we address issues that arise from intrinsic factors in the specific context of nanoparticles of perovskite oxide $\mathrm{La_{0.67}Ca_{0.33}MnO_3}$. In manganites, the existence of competing interactions that arise from orbital, charge and spin degrees of freedom make them an interesting  system. The manganites exhibit a wide spectrum of physical properties leading to rich phase diagrams arising from these competing interactions.\cite{CNRBook}  Thus, the investigation of the effect of size reduction in such a system with a multitude of competing interactions makes it a very interesting study. In particular, it is important to understand how the finite size affects various aspects of the ferromagnetic state including the nature of the paramagnetic  to ferromagnetic transition.

\noindent 
The physical properties of manganites arise from two generic broad classes of competing effects, namely the charge and orbital order that lead to insulating and generally antiferromagnetically (AFM) ordered ground state, and the Double Exchange interactions that lead to metallic and ferromagnetically (FM) ordered ground state. These interactions can be tuned by a number of external parameters and also internal parameters like the carrier concentration, structure and ionic size. The resulting ground state thus depends on the relative strength of these two competing interactions. In the nanomanganites, an additional factor, namely the size, becomes important and can tune these competing interactions and even change the nature of the ground state. In this paper we investigate the effects of size reduction on ferromagnetic nanoparticles of $\mathrm{La_{0.67}Ca_{0.33}MnO_3}$ down to a size of $15$ nm by using magnetic measurements as well as neutron diffraction that gives information both on the magnetic state as well as on the structure. We note that while bulk magnetic measurements have been used in the past to study nanoparticles of manganites, neutron diffraction studies have not been used to study the structure and magnetism in these ferromagnetic nanoparticles. In this investigation we specifically address the following issues: (a) The change in the ferromagnetic transition temperatures on size reduction , (b) the reduction in the spontaneous magnetic moment and (c) the important issue of change in the nature of the ferromagnetic to paramagnetic transition from first order to second order on size reduction. 

\noindent
Out of all the doped manganites, nanoparticles of $\mathrm{La_{0.67}Ca_{0.33}MnO_3}$ (LCMO) are perhaps the most widely studied systems.\cite{TKNath,Lopez,Shantha,Rabinda} The ferromagnetic films of $\mathrm{La_{0.67}Ca_{0.33}MnO_3}$ with thickness down to few tens of nm have also been investigated to look for the effects of size reduction on the ferromagnetic to paramagnetic transition temperature $T_C$.\cite{LCMOFilm} However, quite contradictory results have been published by different groups as to what happens to the ferromagnetic $T_C$ as the particle size (diameter $d$) is reduced.\cite{TKNath,Shantha,Rabinda} The results range from enhancement of $T_C$, reduction of $T_C$ as well as no change in $T_C$ as $d$ is reduced. We re-examine this apparently conflicting issue by making particles of sizes over a large range (down to $15$ nm), that has not been done before, using the same synthesis method. We establish the crucial observation that size reduction leads to a shallow but distinct strengthening of $T_C$ down to a size of nearly $50-75$ nm and then, on further size reduction, we find that the $T_C$ is suppressed as the size is taken down to $15$ nm. Interestingly, in this size range the spontaneous magnetization $M_S$, as measured by the neutron diffraction data show that though a magnetic order still exists in the bulk of the nanocrystals even down to $15$ nm, there is a substantial reduction in the average value of $M_S$. By comparing our data on $\mathrm{La_{0.67}Ca_{0.33}MnO_3}$ with that obtained in nanoparticles of $\mathrm{La_{0.5}Sr_{0.5}CoO_3}$, which is not a Double-Exchange ferromagnet, we establish that the enhancement of $T_C$ seen in the initial stage of size reduction (d $>$ 50 nm) is a special feature of the manganites. 
 
\noindent
Our earlier work on half-doped manganite nanocrystals,\cite{PRB,APL} have shown that the charge and orbitally ordered ground state gets  affected by the structural changes occurring in the system on size reduction. This led us to believe that a study of how the structure of the $\mathrm{La_{0.67}Ca_{0.33}MnO_3}$ nanocrystals evolves on size reduction might well shed some light on its magnetic properties. This paper thus investigates  whether size reduction in $\mathrm{La_{0.67}Ca_{0.33}MnO_3}$ leads to any change in the crystal structure symmetry, or whether there are any changes in the values of the structural parameters. Such detailed structural study on nanocrystals of $\mathrm{La_{0.67}Ca_{0.33}MnO_3}$ have not been done before. 

\noindent  
The issue of the order of paramagnetic to ferromagnetic phase transition in manganites is a topic of considerable interest. Manganite with composition $\mathrm{La_{0.67}Ca_{0.33}MnO_3}$ shows a first order phase transition at the paramagnetic to ferromagnetic transition\cite{polycrystalline1,polycrystalline2, natureoftransition,Yusuf} in the bulk form. This is a special feature of hole doped system $\mathrm{La_{1-x}Ca_{x}MnO_3}$. This happens for $x\approx 0.2-0.4$. For $x \leq 0.2$, the transition is second order with critical exponents close to what one would expect from a 3D-Heisenberg system. For $x \geq 0.4$ the first order transition ends in continuous transition at a tricritical point.\cite{Kim} For other manganites with Sr or Ba substitution instead of Ca,\cite{SrBasubstitution} the paramagnetic to ferromagnetic transition is second order. It thus appears that depending on the bandwidth and/or carrier concentration, the nature of the transition can change. We have investigated the effect of size reduction on the nature of the phase transition in the $\mathrm{La_{0.67}Ca_{0.33}MnO_3}$ system and found that the first order transition changes over to a continuous ($2^{nd}$ order) transition on reducing the size, and the exponents approach a value close to that predicted by the mean field theory. We believe that this particular aspect has not been investigated before in nanoparticles of manganites. 

\noindent
\section{EXPERIMENTAL}
The samples were prepared by a standardized chemical synthesis route and were well characterized using techniques like x-ray diffraction (XRD)\cite{XRD} and imaging tools like transmission electron microscopy (TEM)\cite{TEM} and scanning electron microscopy (SEM).\cite{SEM} The details of the synthesis method can be found in one of our earlier publications.\cite{PRB} We could grow a batch of samples with varying particle diameter (see Table I) using the same synthesis technique. The chemical (cation) stoichiometry was checked using Inductively Coupled Plasma Atomic Emission Spectroscopy (ICPAES) technique, and the oxygen stoichiometry was checked using iodometric titration. All our samples  had slight  oxygen  deficiency  ($\mathrm{La_{0.67}Ca_{0.33}MnO_{3-\epsilon}}$)  with $\epsilon$ positive. $\epsilon \approx 0.02$ for the sample with the smallest particle diameter and it increased somewhat for the bulk sample. Thus the particles with smaller particle diameter have somewhat better oxygen stoichiometry. A list of the samples that were used for the neutron diffraction studies have been shown in Table I. 

\begin{table*}
\caption{\label{tab:table1}Sample ID, average particle diameter ($d$), oxygen deficiency ($\epsilon$), ferromagnetic Curie temperature ($T_C$) and coercive field ($H_C$) of $\mathrm{La_{0.67}Ca_{0.33}MnO_3}$ samples used for neutron diffraction studies.}
\begin{ruledtabular}	
\begin{tabular}{cccccc}
Sample ID& $d$ (nm)& $\epsilon$& $T_C$ from $M$-$T$ measurements (K) & $T_C$ from neutron data (K)& $H_C$ (mT) at T=80K\\
\hline
A&15 $\pm$ 2&0.02&242&234&20\\
B&50 $\pm$ 15&0.04&294&291&23\\
C&300 $\pm$ 40&0.03&277&274&10\\
D&21500 $\pm$ 2000&0.05&265&268&3\\
\end{tabular}
\end{ruledtabular}
\end{table*} 

\noindent 
The Curie temperatures quoted in Table I have been established from magnetization measurements as well as from the neutron data. All the magnetic measurements were carried out using a Vibrating Sample Magnetometer.\cite{VSM} The average particle diameter was estimated both from XRD results (using Williamson Hall plot\cite{WH}) as well as from the imaging tools like Transmission Electron Microscope (TEM) for smaller particles, and Scanning Electron Microscope (SEM) for larger particles. The latter also gave us the size distribution in each sample. The single crystallinity of the particles was also checked using TEM. Fig. 1 gives representative TEM and SEM data for particles with extreme sizes (Sample A and D in Table I). Based on the extensive TEM data we find that the nanoparticles of size range below $100$ nm are single nanocrystals. We will thus use the terms nanoparticles and nanocrystals interchangeably.

\begin{figure}[t]
\begin{center} 
\includegraphics[width=8cm,height=13.3cm]{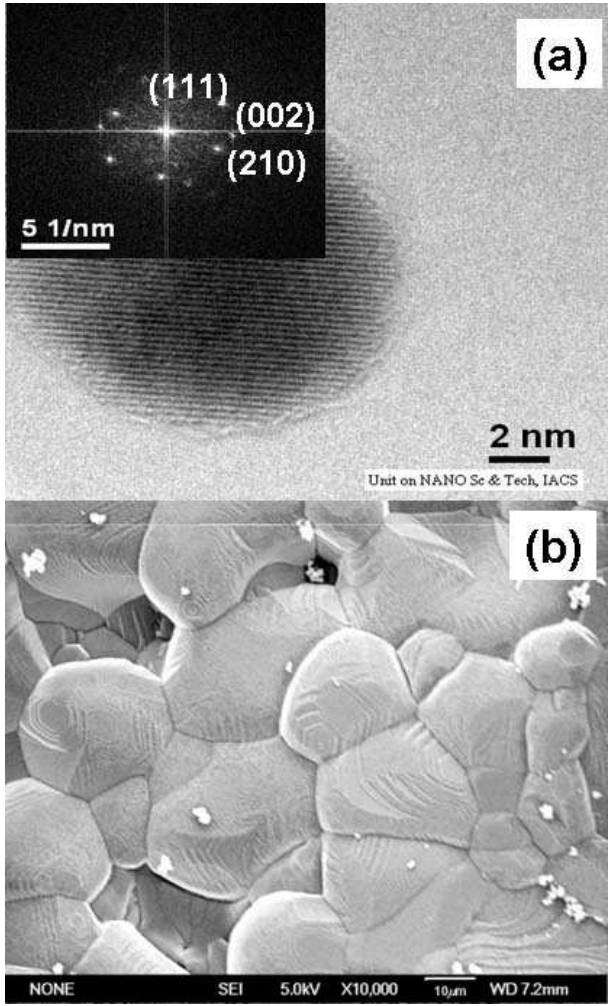}
\end{center}
\caption{(a) TEM image of LCMO-A showing a single nanoparticle of size $\sim$ 15    nm and (b) SEM image of LCMO-D. The inset in (a) shows the electron diffraction pattern taken on LCMO-A.}
\label{Fig1}
\end{figure}

\noindent
The neutron diffraction experiments on these samples were carried out at the Dhruva reactor (beam port T1013) in Bhabha Atomic Research Centre, Mumbai, India. The wavelength used was 1.249 ${\AA}$ and the scattering angular range covered was $2\theta = 5^0$ to $140^0$ with a step size of $0.05^0$. The powders were packed in a Vanadium can of height 35 mm and diameter 6 mm. For low temperature scans, a closed cycle refrigerator was used. This diffractometer has a flux of $8.5 \times 10^5$ $neutrons/cm^2/sec$ and a beam size of 4 cm $\times  1.5$ cm. There are a total of 5 linear position sensitive detectors which allows the Q range to vary from 0.4 ${\AA}^{-1}$ to 9.4 ${\AA}^{-1}$. The Rietveld analyis of the structure (both lattice as well as magnetic structure) was done using the FullProf Suite software.\cite{FP}
 
\noindent 
\section{RESULTS}
\subsection{FERROMAGNETIC $T_C$ AND ITS DEPENDENCE ON SIZE}

\noindent 
The Curie temperatures of the samples were determined from the magnetization data measured at a field of $H$ = 1 mT. The representative data on field cooled (FC) and zero field cooled (ZFC) magnetization ($M$) versus temperature ($T$) curves for the samples are shown in Fig. 2. The samples show a ferromagnetic transition temperature $T_C$ down to the lowest sample size (LCMO-A). Though not shown here, we have obtained well defined hysteresis loops for all the samples, including the sample with the smallest size. At $T=$ 80 K, the coercive field $H_C$ changes from a rather low value of 2 mT for the bulk sample to about 20 mT for the sample A as shown in Table I. The temperature and size variation of the $H_C$ in these materials are interesting but a detailed study of this issue is not within the scope of the paper. 
The magnetic data show that the samples that we are working with remain ferromagnetic even after size reduction of more than three orders of magnitude, although there is a distinct variation of $T_C$ as well as the absolute value of the magnetization with size as described below. 

\begin{figure}[t]
\begin{center} 
\includegraphics[width=8cm,height=7cm]{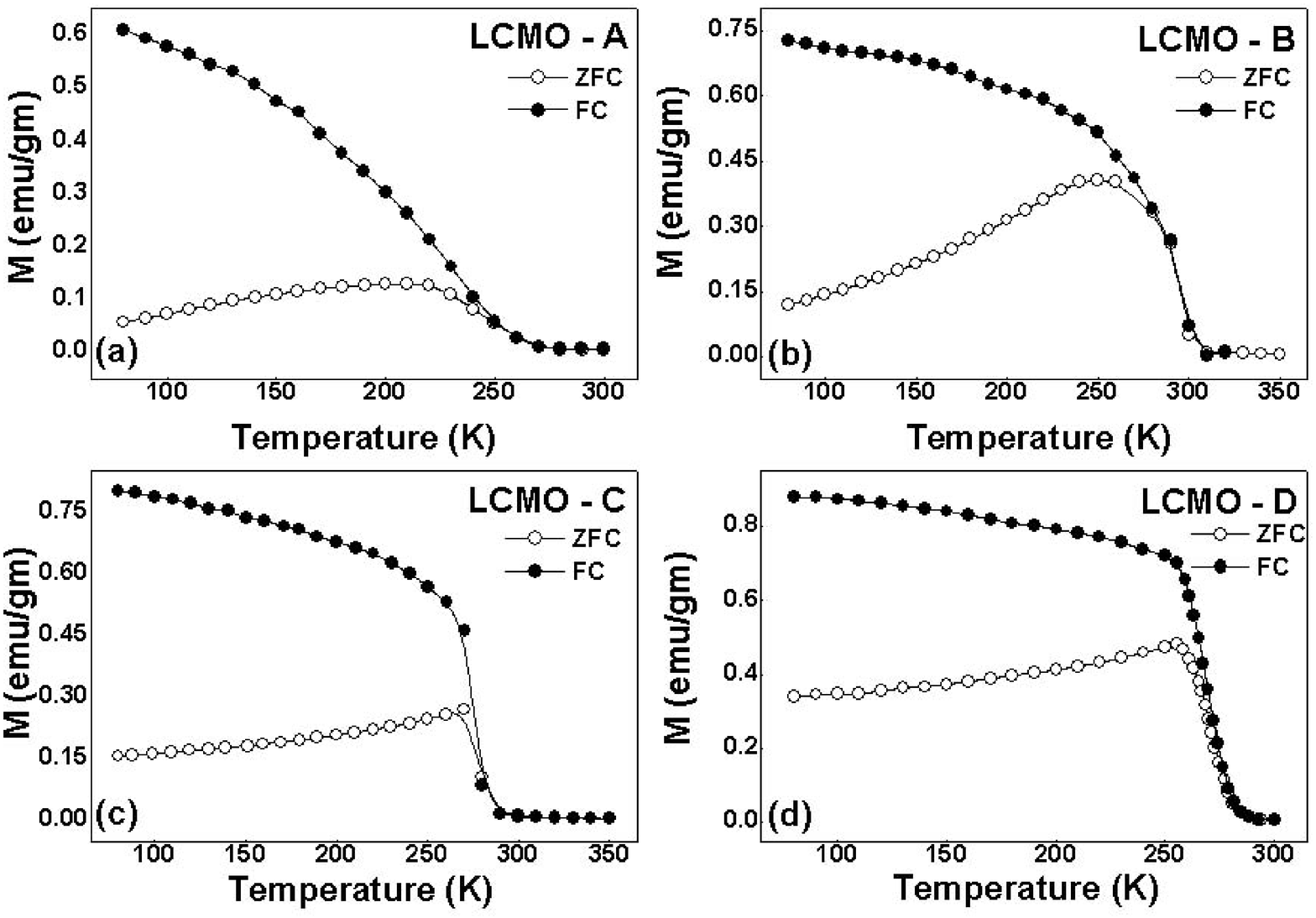}
\end{center}
\caption{Zero field cooled and field cooled M vs T curves for $\mathrm{La_{0.67}Ca_{0.33}MnO_3}$ samples with different average particle diameter.}
\label{Fig2}
\end{figure}

\noindent
In Fig. 3 we show the variation of $T_C$ with particle diameter ($d$). The graph also has data on samples of additional sizes on which magnetization data were taken but neutron data were not taken and they are not shown in Table I. The $T_C$ was identified from the inflection points in the $dM/dT$ versus $T$ plots. In the same figure, we also plot the $T_C$ values as obtained from the neutron diffraction data (we elaborate on the neutron diffraction measurements in the next section). The data obtained from the two methods are very close except for the smallest sample, where the neutron data shows somewhat less ($\approx 3\%$) $T_C$. This is understandable because the magnetic transition in the smaller size nanoparticles are broader which does introduce some uncertainty in its determination. Also, in the neutron data a clear $T_C$ can be seen only when a long-range magnetic correlation builds up so that the intensity that adds on to a structural peak can be clearly distinguished. The identification of a clear $T_C$ by the neutron data thus is significant because it denotes onset of bulk ferromagnetic order in appreciable part of the sample. We also note that no past studies have settled this issue whether a long-range ferromagnetic order indeed exists in such small manganite nanoparticles of size as small as $15$ nm. We discuss this issue again later on.

\noindent 
We note here that by varying the particle size over such a wide range we find that our samples show a non-monotonic variation of the Curie temperature with $d$. This is unlike earlier reports of magnetic studies on $\mathrm{La_{0.67}Ca_{0.33}MnO_3}$ nanocrystals (done with limited range of particle size) which reported a monotonic variation or no variation of $T_C$ with particle diameter.\cite{Shantha,TKNath} This result, i.e., a non-monotonic variation of $T_C$  with size, is thus new. We will see below that it points to an interesting physics where we have a likely combination of finite-size effect and effects due to variation of the bandwidth working in tandem. The size where the turn around in $T_C$ occurs is for $d \approx$ 50 nm, and we will see that there is a reduction of the magnetic moment (as determined from the neutron data) in this size range. 

\noindent
 We observe that the region of turn around in $T_C$ is somewhat sensitive to the method of sample preparation/heat treatment temperatures and variations in $T_C$ of  $\approx \pm 5\%$ can be seen in the size range $d\approx 70-100$ nm (where the turn around occurs) depending on the sample preparation conditions. The sample preparation condition changes the size distribution even if the average size is the same. Since $T_C$ has a non-monotonic dependence on $d$, this makes the average value of the observed $T_C$ rather sensitive to the exact size distribution leading to variations in $T_C$ for the same average size $d$.
 
\begin{figure}[t]
\begin{center}
\includegraphics[width=8cm,height=7cm]{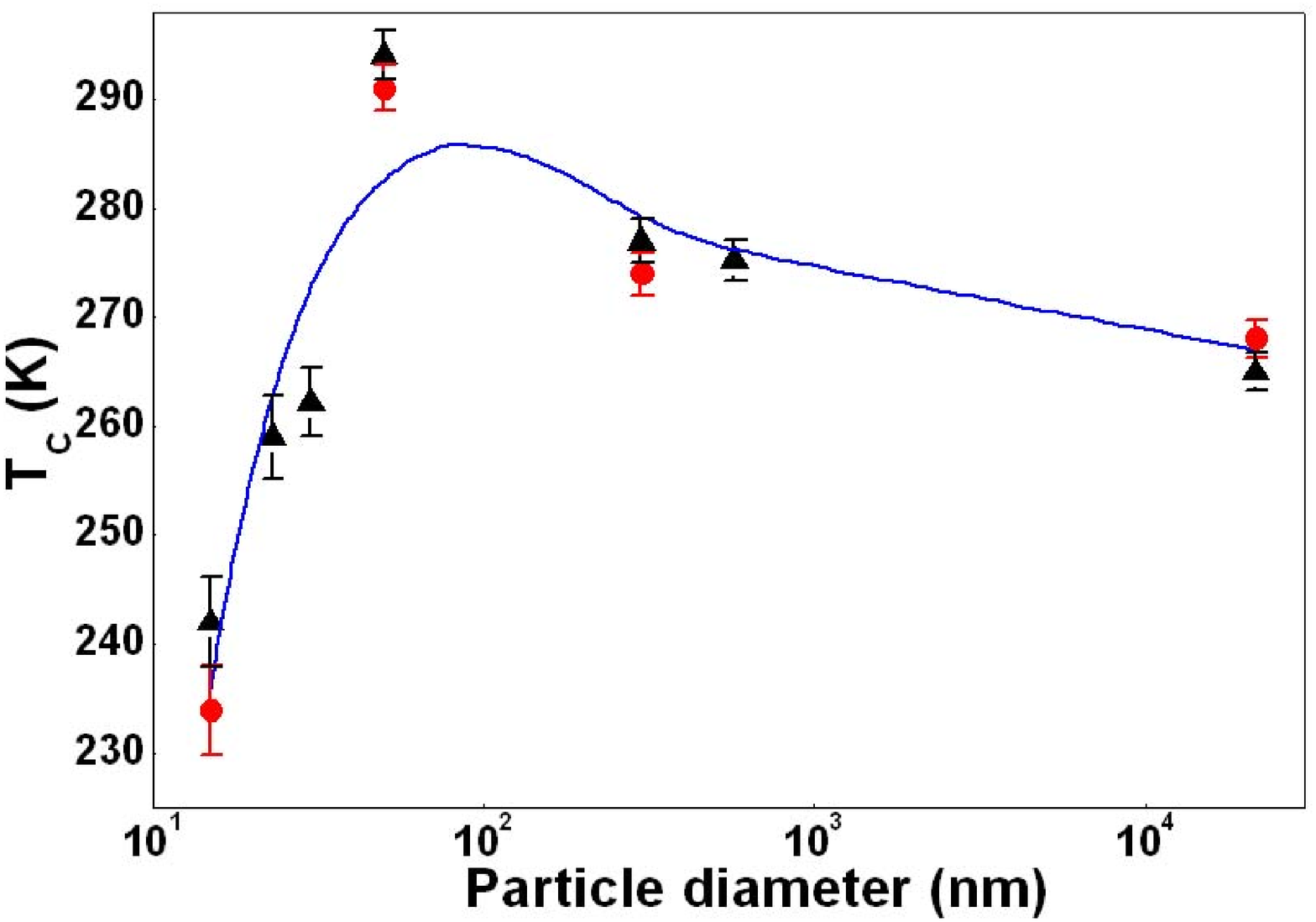}
\end{center}
\caption{Variation of $T_C$ with $d$ in $\mathrm{La_{0.67}Ca_{0.33}MnO_3}$ nanoparticles. $T_C$ obtained from the magnetization data (black triangles) as well as from the neutron data (red circles) are shown. The line is a fit to the data using equation 5 (see Discussion section).}
\label{Fig3}
\end{figure}

\noindent
\subsection{LATTICE AND MAGNETIC STRUCTURES FROM NEUTRON DATA} 
An important issue that the paper probes is whether there is indeed a change in the lattice as well as the magnetic structure (including the spontaneous magnetization) on size reduction and whether the observed changes in the magnetic properties can have a structural basis. This is relevant in the light of our earlier studies\cite{PRB,APL} which have shown that the crystal structure is strongly related to the orbitally ordered ground state of these systems in the regime of carrier concentration where there is charge ordering. To this purpose, neutron diffraction measurements were done on four representative samples (listed in Table I). This gave us a dual advantage of being able to probe the lattice as well as the magnetic structure. The data were taken at different temperatures from 20 K to 320 K. A detailed two-phase Rietveld refinement (with a structural phase and a magnetic phase) was done to extract the various structural and magnetic parameters. The space group $Pnma$ was used for both the phases. The line profiles were modeled using a pseudo-Voigt profile shape function. As an example, in Fig. 4, we show the neutron diffraction patterns for LCMO-A (along with the fits) taken at (a) $T$ = 300 K and (b) $T$ = 20 K. In the corresponding insets, we have shown the expanded regions between $2\theta = 16^0$ and $27.5^0$ to show the two Bragg peaks, ((101) and (200)), where the magnetic contribution has been observed. $\mathrm{La_{0.67}Ca_{0.33}MnO_3}$ being ferromagnetic, the magnetic contribution occurs at the same $2\theta$ values as the structural peaks. The enhancement of the Bragg peak intensity with decreasing temperature (as can be seen clearly from a comparison of the intensities in the insets of Fig. 4) confirms the presence of ferromagnetic ordering in the sample with the lowest particle size.

\begin{figure}[t]
\begin{center}
\includegraphics[width=8cm,height=7cm]{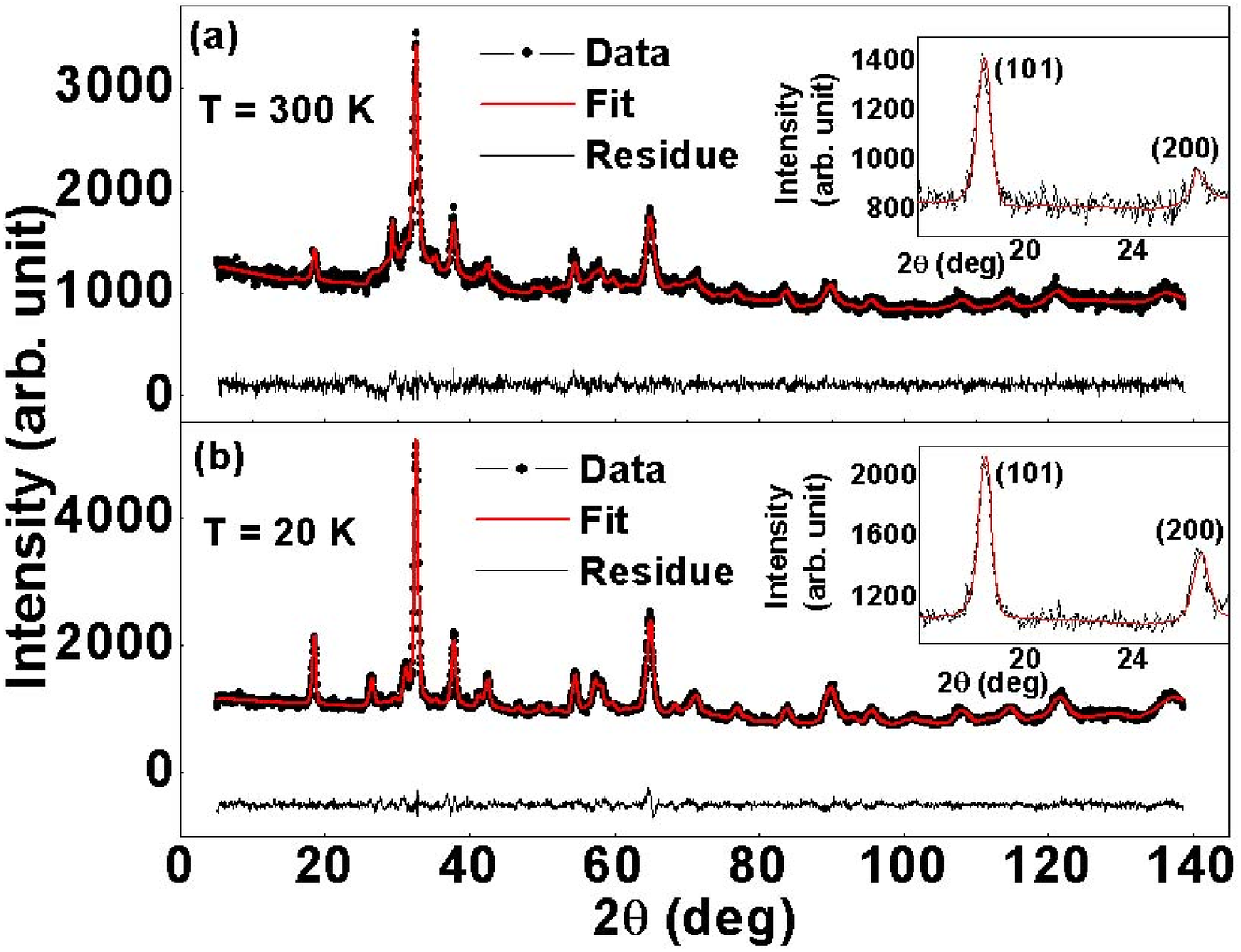}
\end{center}
\caption{Neutron diffraction patterns for the nanoparticles with size $15$ nm (LCMO-A) taken at (a) $T$ = 300 K and (b) $T$ = 20 K. The insets show the expanded regions between $2\theta = 16^0$ and $27.5^0$. The line through the data points are the results of profile fitting. The residues are shown at the bottom.}
\label{Fig4}
\end{figure}

\noindent
The Rietveld refinement was used to obtain the lattice parameters as well as the bond angles and bond lengths of the samples. The temperature evolution of the lattice parameters shows that there are no large lattice distortions at the transitions but the lattice shows a small contraction at the magnetic transition. This is similar to that found in manganites like $\mathrm{La_{0.7}Sr_{0.3}MnO_3}$.\cite{LSMO} The coefficient of thermal expansion ($\alpha$) calculated from the temperature dependence of the cell volume shows that the value of $\alpha$ remains almost unchanged at $\approx 1 \times 10^{-5}/K$ despite a decrease in the particle diameter by three orders.  The evolution of the lattice parameters and the cell volume with size at two temperatures ($T$ = 20 K and $T$ = 300 K) is shown in Fig. 5. There is not much change in the cell volume except for the smallest size sample where it shows a small increase.  This is in sharp contrast to the half doped $\mathrm{La_{0.5}Ca_{0.5}MnO_3}$\cite{PRB,APL} where size reduction leads to a distinct contraction in the cell volume.

\begin{figure}[t]
\begin{center}
\includegraphics[width=8cm,height=7cm]{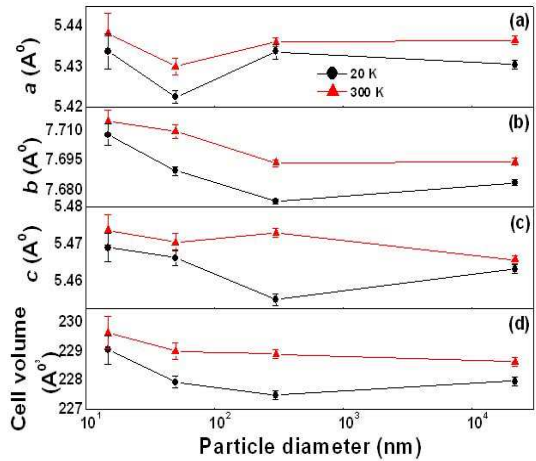}
\end{center}
\caption{Variation of (a) {\it a}, (b) {\it b}, (c) {\it c} and (d) unit cell volume with particle diameter at $T$ = 20 K and $T$ = 300 K for $\mathrm{La_{0.67}Ca_{0.33}MnO_3}$.}
\label{Fig5}
\end{figure}

\noindent 
The analysis of the neutron data shows that a clear change, although small, occurs in the $MnO_6$ octahedron on size reduction as revealed through the Mn-O-Mn bond angles and Mn-O bond lengths. These two parameters have a direct role in determination of the overlap integral and thus on the bandwidth $W$. In Fig. 6 we show the temperature dependence of the two Mn-O-Mn bond angles for the samples studied. Bond angles $Mn-O_{1}-Mn$ and $Mn-O_{2}-Mn$ refer to the apical ($O_1$) and equatorial ($O_2$) oxygen atoms. In Fig. 7, the temperature variations of the two bond lengths for the samples are shown. The size reduction thus leads to a gradual compaction of the Mn-O bond lengths and Mn-O-Mn bond angles move closer to 180$^0$ for the apical oxygen. Both these are expected to enhance the band width $W$. 

\begin{figure}[t]
\begin{center}
\includegraphics[width=8cm,height=7cm]{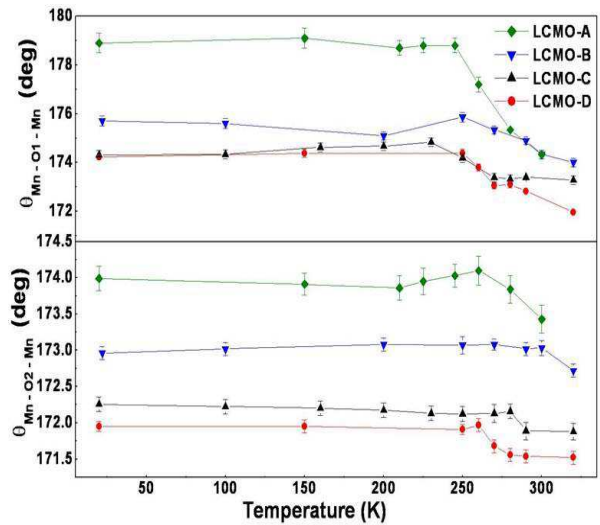}
\end{center}
\caption{Variation of the Mn-O-Mn bond angle with temperature for $\mathrm{La_{0.67}Ca_{0.33}MnO_3}$ with varying particle diameters.}
\label{Fig6}
\end{figure}

\begin{figure}[t]
\begin{center}
\includegraphics[width=8cm,height=7cm]{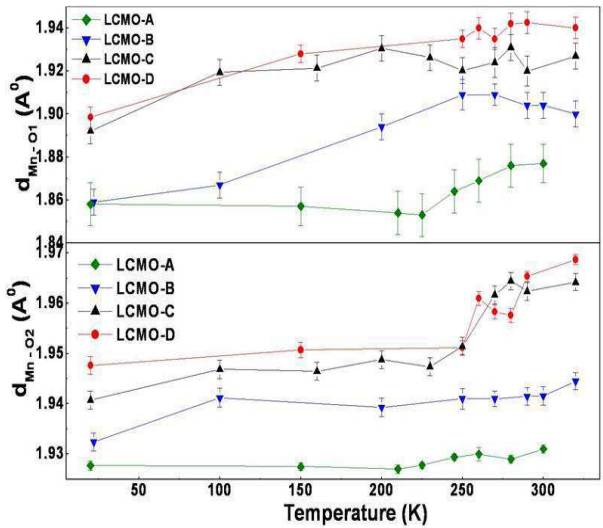}
\end{center}
\caption{Variation of the Mn-O bond length with temperature for $\mathrm{La_{0.67}Ca_{0.33}MnO_3}$ with varying particle diameters.}
\label{Fig7}
\end{figure} 

\noindent 
The unit cell of the manganites is often distorted from the cubic structure. The Jahn-Teller distortion around the $Mn^{3+}$ ions leads to orthorhombic distortions. The reduction of the orthorhombic distortion strengthens the double-exchange interaction, enhances the band width and strengthens the ferromagnetic state. Ferromagnetic composition $\mathrm{La_{0.67}Ca_{0.33}MnO_3}$ has low orthorhombic strains. The orthorhombic distortion can be quantified by   orthorhombic strains ($O_{S_{\parallel}}$ and $O_{S_{\perp}}$) which give a measure of the deviation of the unit cell from the perfect cubic structure. We define the orthorhombic strains as $O_{S_{\parallel}} = 2(c-a)/(c+a)$ (giving the strain in the $ac$ plane), and $O_{S_{\perp}} = 2(a+c-b\sqrt{2})/(a+c+b\sqrt{2})$ (giving the strain along the $b$ axis with respect to the $ac$ plane). In Fig. 8 we show the evolution of the orthorhombic strains as a function of temperatures for the four samples. We find that there is a small but very discernible reduction in the $O_{S_{\perp}}$ strains on size reduction. (Note: The strains are small and have been calculated by taking differences of two relatively large but nearly equal quantities. This enhances the error bar on the strain data.) At this stage it will be worthwhile to compare the structural changes with compositions that are more distorted. In half-doped $\mathrm{La_{0.5}Ca_{0.5}MnO_3}$, due to Jahn-Teller distortion, the onset of charge and orbital ordering leads to strong orthorhombic strain which gets qualitatively affected on size reduction. In the ferromagnetic $\mathrm{La_{0.67}Ca_{0.33}MnO_3}$, the orthorhombic strain is low to start with and thus there is no large effect on this strain on size reduction. Thus, there is a fundamental difference in the structure of nanocrystals of these compositions that show different ground states.  

\begin{figure}[t]
\begin{center}
\includegraphics[width=8cm,height=7cm]{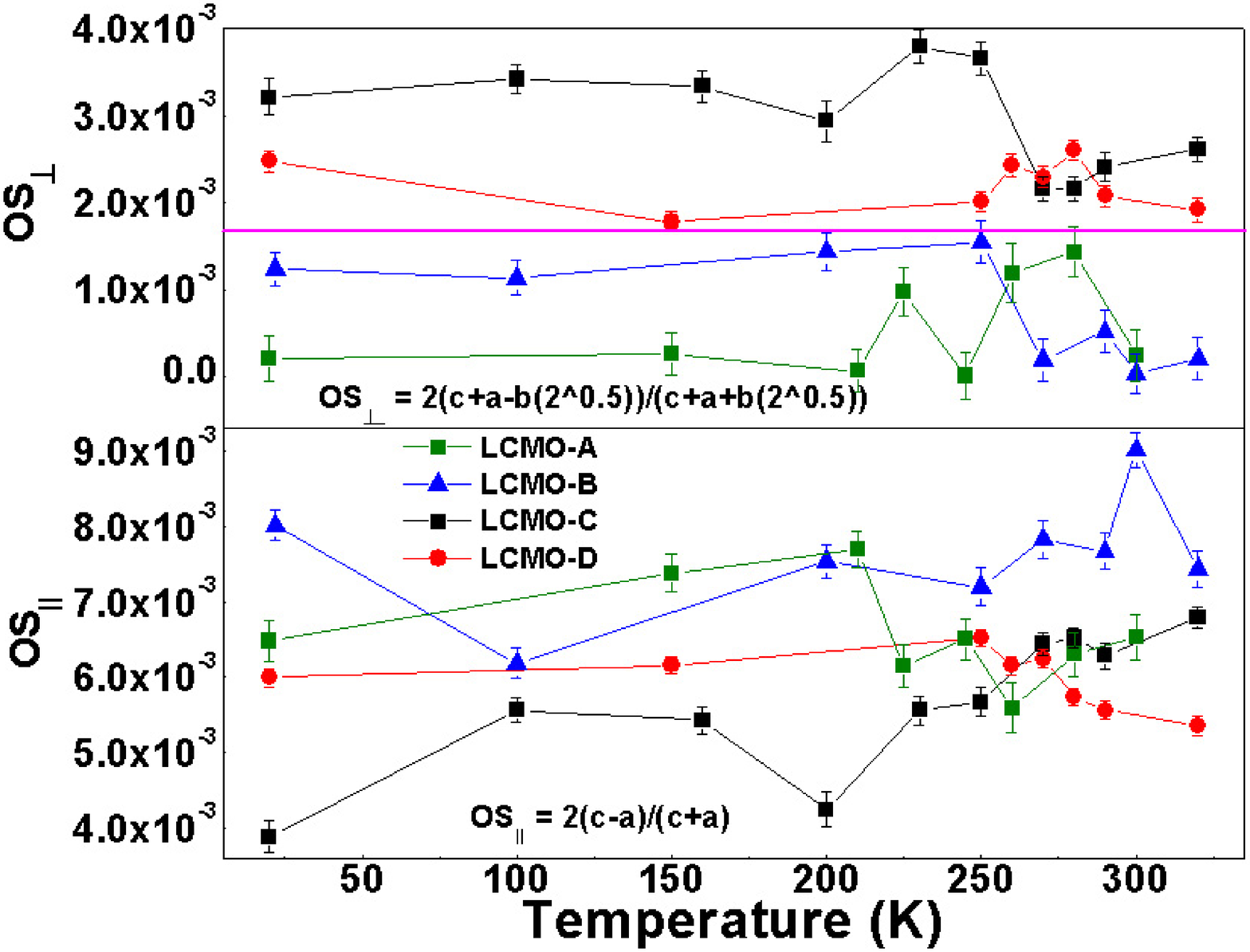}
\end{center}
\caption{Variation of $O_{S_{\parallel}}$ (lower panel) and $O_{S_{\perp}}$ (upper panel) with temperature for $\mathrm{La_{0.67}Ca_{0.33}MnO_3}$ with varying particle diameters. The pink line in the upper panel shows a demarcation between the larger and smaller diameter particles.}
\label{Fig8}
\end{figure}

 \noindent
The analysis of structural data as obtained from the neutron data, clearly shows that the size reduction even down to a diameter of $15$ nm retains the basic perovskite structure and does not distort it in any significant way. The small yet clear change that occurs is that the $MnO_6$ octahedra becomes somewhat more compact and the Mn-O-Mn angle, both for the apical and equatorial oxygens, move closer to 180$^0$. Such a change, as we will see below, can lead to an enhancement of the bandwidth. 

\subsection{VARIATION OF SPONTANEOUS MAGNETIZATION $M_S$ WITH $T$ and $d$}

\noindent 
Neutron data allow us to obtain the spontaneous magnetization ($M_S$), which is measured at zero applied field. This allows us to answer a fundamental question whether the manganite nanoparticles have a spontaneous magnetization that arises from a long-range ferromagnetic order. In Fig. 9 we show the variation of $M_S$ as a function of $T$ for the samples studied. The variation of $M_S$ at $T=20$ K as a function of the diameter $d$ is shown in Fig. 10. The value of $M_S$ remains close to the bulk value down to $d=50$ nm and then shows a down turn as the size is reduced further. One may note from Fig. 3. that the $T_C$ shows an increase till the size is reduced to this value ($d=50$ nm) and then shows a sharp drop when the size is reduced further. It is to be noted however, that even in the smallest sample ($d=15$ nm), there is a ferromagnetic order and a finite spontaneous magnetization $M_S$ though it is reduced considerably from the bulk value.   

\begin{figure}[t]
\begin{center}
\includegraphics[width=8cm,height=7cm]{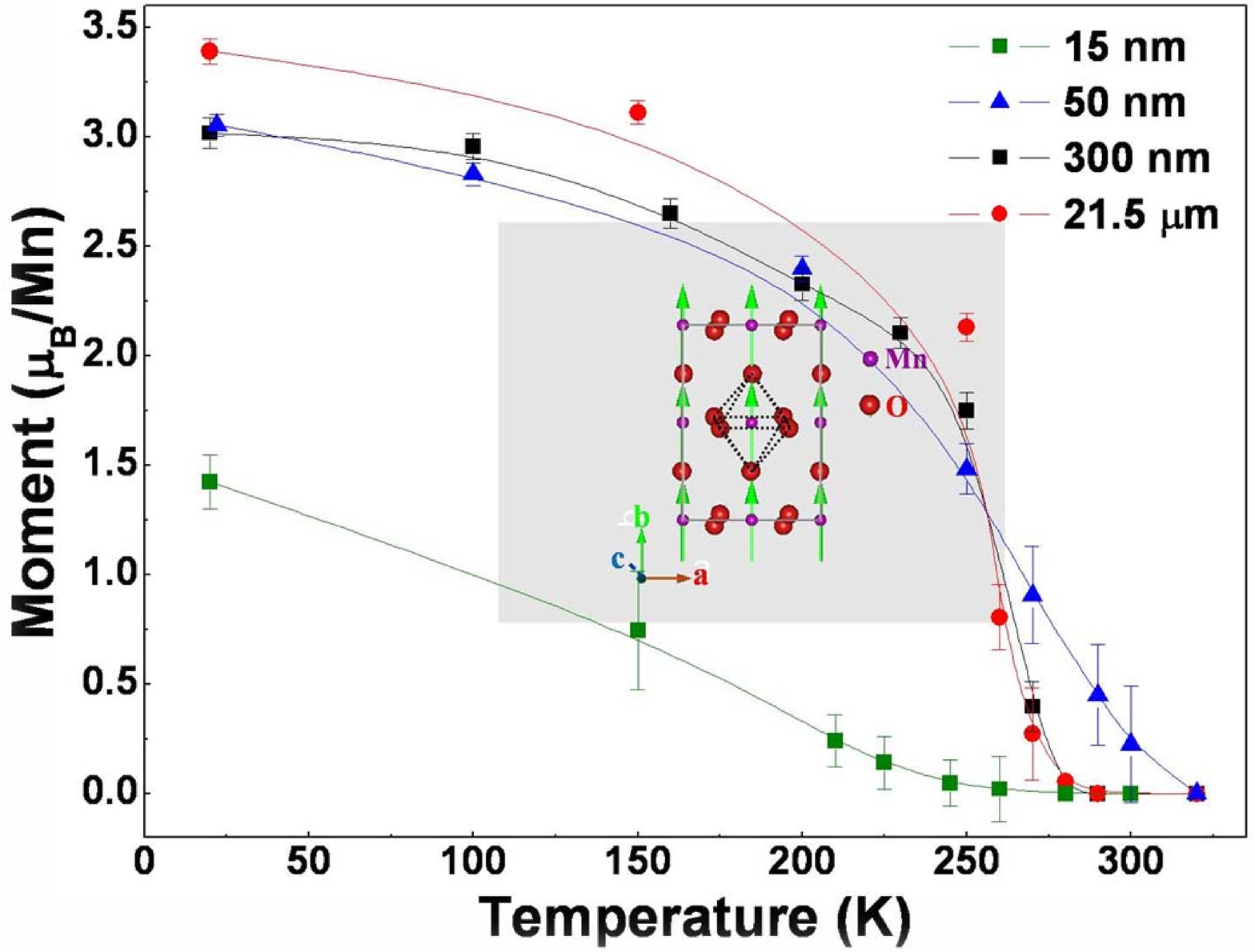}
\end{center}
\caption{Variation of the spontaneous moment with temperature for LCMO-A, LCMO-B, LCMO-C and LCMO-D. The inset shows a schematic representation of the magnetic structure with all the moments aligned along the $b$-axis.}
\label{Fig9}
\end{figure} 

\begin{figure}[t]
\begin{center}
\includegraphics[width=8cm,height=7cm]{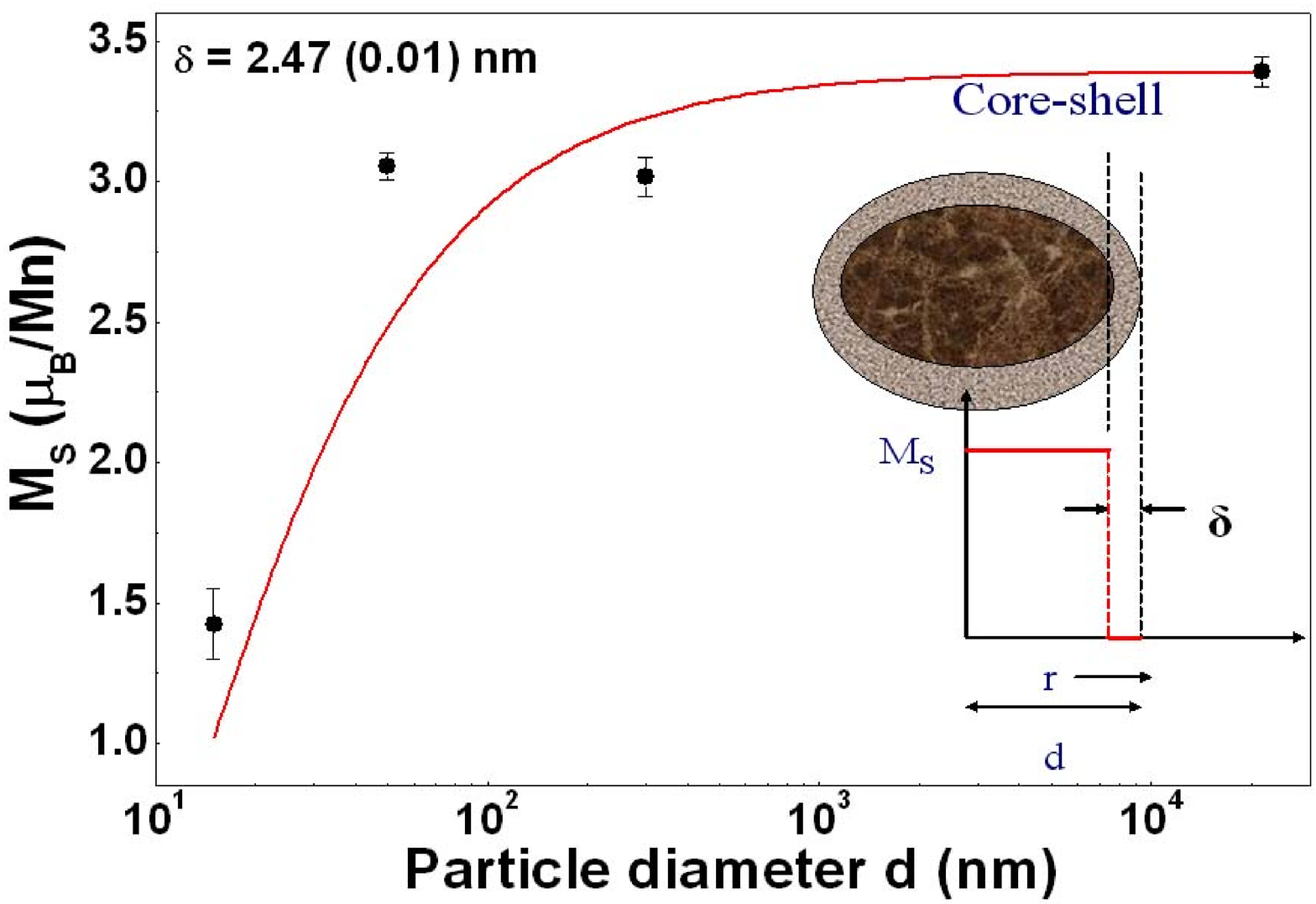}
\end{center}
\caption{Variation of spontaneous moment at T = 20 K as a function of the average particle diameter ($d$). The red line is obtained from the core-shell model using Eqn. 1 (see Discussion section). The inset shows the schematic representation of the core-shell model for the nanoparticles.}
\label{Fig10}
\end{figure} 

\noindent
\subsection{NATURE OF THE MAGNETIC PHASE TRANSITION IN MANGANITE NANOCRYSTALS}
As a part of the investigation on ferromagnetic nanocrystals of $\mathrm{La_{0.67}Ca_{0.33}MnO_3}$ we have investigated what happens to the first order ferromagnetic to paramagnetic phase transition on size reduction.  The nature of the magnetic transition can be obtained from the slope of isotherm plots of $M^2$ vs. $H/M$, $M$ being the experimentally observed magnetization and $H$ the magnetic field. This criterion of deciding the nature of the transition is generally referred to as the Banerjee criterion\cite{banerjee} and has been widely used to experimentally determine the order of the magnetic phase transition. Briefly, a positive or a negative slope of the experimental $H/M$ vs. $M^2$ curve indicates a second order or a first order transition respectively. In the context of materials undergoing second order transition such a plot is called an Arrott Plot. For materials showing first order magnetic transition an Arrott Plot of the type seen for a second order transition will not occur. We have applied this criterion to the $\mathrm{La_{0.67}Ca_{0.33}MnO_3}$ samples with two widely different particle diameters but with a $T_C$ that are rather close. Interestingly, we found striking differences in the magnetic behavior of the two samples around $T_C$. This crossover of the first order transition to the second order transition on size reduction has not been reported before for manganites. There are, however, reports of change in the nature of transition from first order to second order in the AFM transition in MnO nanoparticles supported in porous matrix with sizes down to $14$ nm.\cite{MnO} 

\noindent
Here we note that the presence or absence of hysteresis in $M$ vs $T$ can also be used to establish the nature of the phase transition (presence of hysteresis in Field Cooled Cooling (FCC) and Field Cooled Warming (FCW) data in the case of first order phase transition and absence of the same in the case of second order phase transition). However, in the specific case of bulk $\mathrm{La_{0.67}Ca_{0.33}MnO_3}$, the phase transition appears to be weakly first order. The thermal hysteresis observed in FCC and FCW data is very small ($<$ 2 K).\cite{TangJMM} As a result, using the hysteresis criteria appears to be difficult. The Banerjee criterion provides a more striking and straight forward way to distinguish between samples showing a first order transition, and those showing a second order transition, since the slope of the $M^2$ vs. $H/M$ curves change sign. In addition, exploring the nature of the phase transition through the Banerjee criterion allows us the advantage of determining the critical exponents for the sample exhibiting a second order phase trasntion.

\noindent 
The choice of the samples were guided by the variation of $T_C$ with $d$ as shown in Fig. 3. It can be seen that samples with $d\approx 30$ nm are expected to have $T_C$ close to the bulk value. The investigation was thus carried out on a sample with average $d\approx 23$ nm (referred as SAMPLE-E) which has a $T_{C}\approx 260$ K, which is close to the $T_{C}=265$ K of the "bulk sample" (SAMPLE-D) that has an average size of $d=21.5 \mu$m, which is three orders larger. 

\noindent
We have measured the initial magnetization isotherms in the close vicinity ($\pm 10$ K) of $T_C$. The critical exponents for the continuous transition has been measured for $\epsilon = |T-T_C|/T_{C} < 4 \times 10^{-2}$. Before each run, samples were heated above their $T_C$ and cooled to the measuring temperature under zero field in order to ensure a perfect demagnetization of the samples. In Fig. 11, we plot $M^2$ versus $H/M$ isotherms between 260 K and 280 K for the bulk sample (SAMPLE-D). The deviation from the normal behavior expected from a second order transition is very clear and this deviation is a signature of the first order transition. It is clear that the isotherms present negative slopes in some parts which, according to the criterion used here, is an indication of the first order character of the transition.  The observation of a first order transition in the bulk sample is in agreement with past studies.\cite{polycrystalline1,polycrystalline2,Yusuf}

\begin{figure}[t]
\begin{center}
\includegraphics[width=8cm,height=7cm]{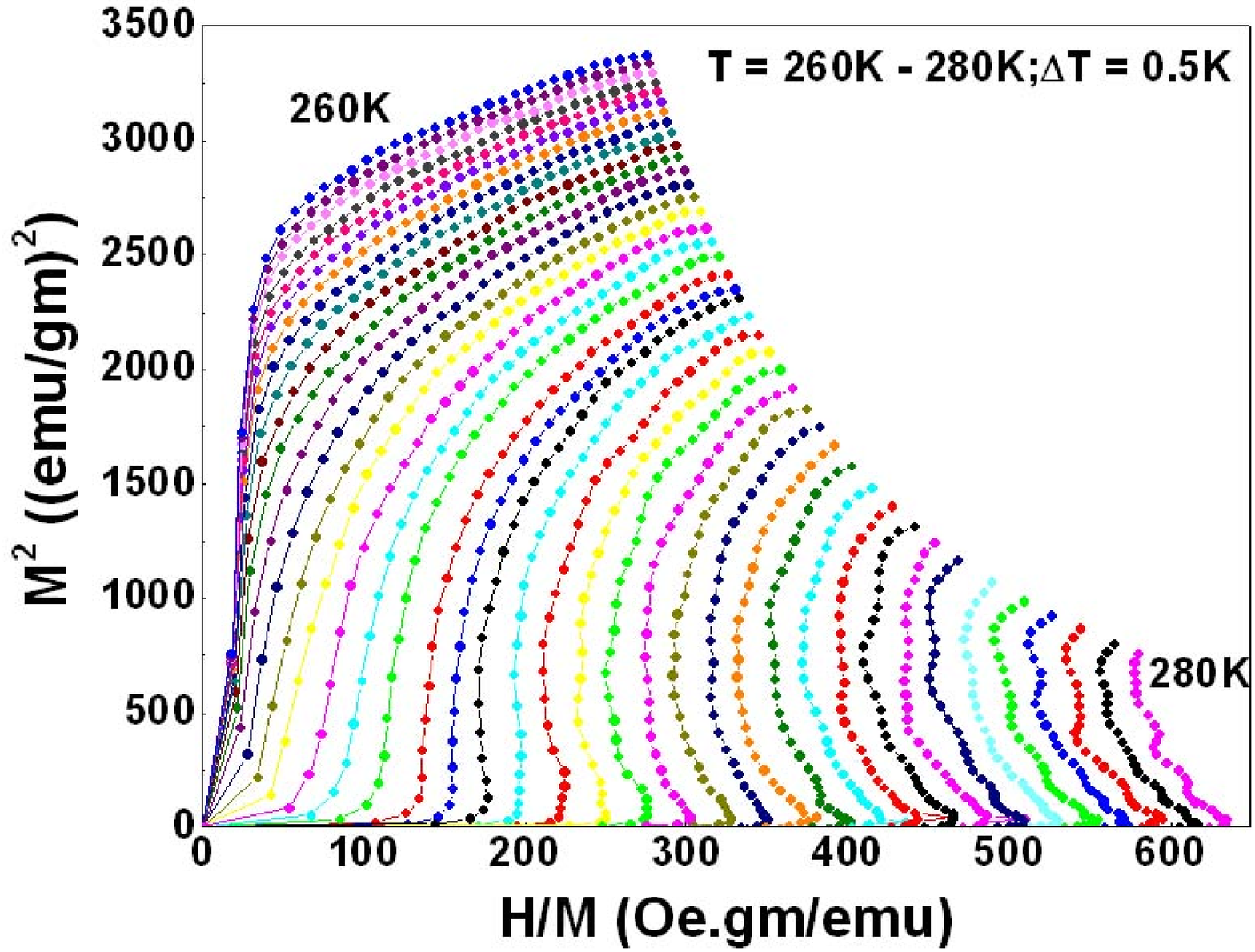}
\end{center}
\caption{$M^2$ vs. $H/M$ plots for bulk LCMO (SAMPLE-D) with $d=21.5 \mu$m.}
\label{Fig11}
\end{figure}

\noindent 
The magnetization isotherms for the SAMPLE-E plotted as Arrott plots are shown in Fig. 12. The isotherms were recorded for the temperature range 250 K to 270 K. It is seen that the isotherms do not display the anomalous change of slope as seen in the bulk sample. Here, we find a positive slope throughout the range of $M^2$. Nanoparticles of $\mathrm{La_{0.67}Ca_{0.33}MnO_3}$ thus, show a second order magnetic phase transition at $T_C$. The exact values of the critical exponents $\beta$ and $\gamma$ and the exact Curie temperature $T_C$ were determined from a modified Arrott plot by taking $\beta$, $\gamma$ and $T_C$ as parameters to be obtained from the fit. (The plot has not been shown to avoid duplication). The exact values of the exponents are $\beta=0.47\pm 0.01$ and $\gamma=1.06\pm 0.03$. One can also obtain $\delta$ directly from the plot of $M$ vs $H$ at $T_{C}$. From such a plot we obtain $\delta=3.10 \pm 0.13$. The mean field value of the exponent $\delta$ can also be derived using $\delta = 1 + \frac{\gamma}{\beta}$. We find $\delta = 3.26 \pm 0.16$. A comparison of the critical exponents show that the magnetic transition is not only second order but the exponents are closer to the mean field values: $\beta= 0.5$, $\gamma= 1$ and $\delta = 3$ than what one would expect them to be for a 3D-Heisenberg system ($\beta = 0.365$, $\gamma = 1.386$  and $\delta = 4.80$). 

\begin{figure}[t]
\begin{center}
\includegraphics[width=8cm,height=7cm]{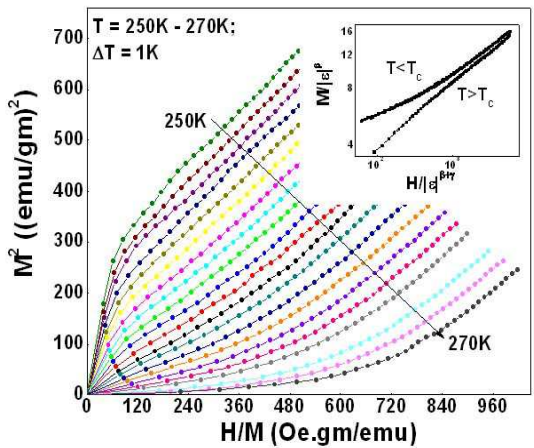}
\end{center}
\caption{Arrott plots for nanocrystals of LCMO (SAMPLE-E) with $d=23$ nm. The inset shows the scaling plots.}
\label{Fig12}
\end{figure}

\noindent
The critical exponents of the transition can be equivalently determined from scaling plots of the form $M/|\epsilon|^\beta = f_\pm(H/|\epsilon|^{\gamma+\beta})$, where, $\epsilon = |T-T_C|/T_C$, $f_\pm$ is a scaling function, and the plus and minus signs correspond to the ferromagnetic and paramagnetic regions respectively. By appropriate selection of the parameters $T_C$, $\beta$ and $\gamma$, the data should collapse on two different branches for $T > T_C$ and $T < T_C$. We construct scaling plots (shown in the inset of Fig. 12) to prove the validity of our choice of $\beta$, $\gamma$ and $T_C$. A convincing scaling of the data points on the two branches of the scaling function $f_\pm$ can be seen.

\noindent
We have thus, observed a very clear change in the nature of the paramagnetic to ferromagnetic transition in $\mathrm{La_{0.67}Ca_{0.33}MnO_3}$ from first order to second order when the particle diameter is brought down from bulk to few tens of nanometers range. We note that the change in the nature of the phase transition by changing the size has not been reported before. Though we have not investigated the change in the nature of the transition by varying the particle sizes in close steps, the preliminary data that are available shows that the cross-over may occur for average sizes less than $200$ nm.

\noindent
\section{DISCUSSIONS}
The results presented above show that there are important and subtle changes that occur in the magnetic properties on size reduction in manganites. While the ferromagnetic state is retained down to a size of $d=$15 nm, the magnetization is suppressed beyond a size of $d\approx$ 50 nm. The observation that the ferromagnetic $T_C$ shows a non-monotonic dependence on the size is also interesting. While the suppression of $T_C$ on size reduction may be expected due to such phenomena as finite size effect,\cite{LCMOFilm} the initial enhancement of $T_C$ when the size is reduced to $\approx 50$ nm is unusual. The size reduction also changes the nature of the ferromagnetic transition from a first order to a second order with critical exponents that are close to the mean field values. The crystal structure as obtained from the neutron data shows that there are small yet definite changes on size reduction. These changes are much smaller than the qualitative changes seen in the half-doped composition that has strong orthorhombic strain. The basic perovskite structure of the bulk manganite is preserved although there is a clear compaction of the $MnO_6$ octahedra and the orthorhombic distortion $O_{S_{\perp}}$ is somewhat reduced. We will discuss below whether these small but definite changes on size reduction have any effect on the magnetic properties. In the following subsections we discuss the physical implications of the main observations. 

\noindent
\subsection{DEPENDENCE OF SPONTANEOUS MAGNETIZATION ($M_S$) ON SIZE ($d$)}
The size dependence of $M_S$ has been shown in Fig. 10. Reduction in the saturation magnetization with size reduction has been seen in different types of magnetic nanoparticles before\cite{MN2} including manganite nanoparticles.\cite{Lopez} Our observation however, is made on spontaneous magnetization (obtained with no applied field) as obtained from the neutron data. To explain the reduction of $M_S$ with $d$ we propose a simple core-shell model. This is similar to that proposed for explaining reduction in saturation magnetization in magnetic nanoparticles in general.\cite{coreshell} The magnetic particles have ordered cores that can show long-range ferromagnetic order. We propose that the disordered shell (of thickness $\delta$) does not have any spontaneous magnetization and is like a magnetic dead layer. There are recent TEM studies\cite{NM9} that show that surface of manganite nanowires have a layer of thickness $\approx 2$ nm that are disordered. A close inspection of the TEM data of the type shown in Fig. 1, shows that in most of our nanoparticles there are layers of size $\sim$ $1-2$ nm where no clear lattice fringes are seen. This would thus indicate the existence of structurally disordered regions at the surface. It may happen that the dead layer have spins that are aligned in random directions due to randomness of local anisotropy field. These regions however, lack any long-range order and it is expected that the spontaneous magnetization $M_{S}\rightarrow 0$ in this region. The schematic of the core-shell model is shown in inset  of Fig. 10. If the shell thickness $\delta$ is indeed small, the larger particles with $d >> \delta$ will not be affected, while for smaller particles, $\delta$, though small, can be comparable to $d$ and the measured magnetization will thus be affected. The measured $M_S$ is proportional to the volume fraction of the core that carries the spontaneous magnetization i.e. $M_S = M_{0S}({\frac{\frac{d}{2}-\delta}{\frac{d}{2}}})^3$, where $M_{0S}$ is the spontaneous magnetization for a bulk sample that has long-range magnetic order. Expanding the cube and ignoring higher powers of $\frac{\delta}{d}$, we obtain: 
\begin{equation} 
M_S = M_{0S}(1-\frac{6\delta}{d})
\end{equation} 
Eqn. 1 can be used to find the thickness $\delta$ of the shell. Generally it is expected that $\delta$ may vary for particles of different sizes. However, we find that $\delta$ is rather similar for all the sizes and we can fit the data to Eqn. 1 with a more or less constant value of the shell thickness of $\delta \approx 2.47$ nm as shown by the red line in Fig. 10. The increasing importance of the disordered shell for the smaller diameter particles becomes apparent when we compare $\delta$ to the diameter of the particles ($d$). The rapid reduction of the $M_S$ for small $d$ can thus be understood. From the value of $\delta$ it can be seen that for nanocrystals with $d < 10$ nm the $M_{S}\rightarrow 0$ and it is likely that there will be no spontaneous moment for particles with $d\approx 5$ nm or less.  

\noindent 
The thickness of the surface layer $\delta$, estimated by the size dependence of the magnetization, is similar to that observed from direct TEM imaging.\cite{NM9} This is also in agreement with estimation from local conductance mapping in nanostructured films of manganites as done by a Scanning Tunneling Microscope.\cite{Sohini} In this context we note that the estimation of the value of $\delta$ from spontaneous magnetization $M_S$ as measured from neutron data can be different from that estimated from the magnetization data that is done with an applied field. This is because Eqn. 1 assumes that $M_S=0$ for the shell layer at the surface. However, the surface layer may lack long-range order, which may arise due to freezing of spins at random directions. In this case also $M_S$ will be zero but it may show a finite value of magnetization when a field is applied to measure the magnetization. We note that there is indeed a controversy about the magnetic nature of the surface layer which in some materials can show a low temperature spin freezing which can be interpreted as a spin glass like transition.\cite{NM7} In absence of a detailed study of the surface spin we cannot comment on the exact spin structure at the surface. However, whatever be the nature of the disordered spins at the surface, it will lack long-range order and will have a zero value for the spontaneous magnetization.  Thus, estimating the thickness of the "dead" surface layer from the spontaneous magnetization $M_S$ does not suffer from the ambiguity that may occur in estimates based on saturation magnetization data. (Note: the existence of the disorder in the surface spins are often obtained from the high field susceptibility. Measurements of high field susceptibility in fields upto 1.5 T, which is much larger than the field for technical saturation, showed that there is no large change in the high field susceptibility on size reduction.)

\noindent
Here, we would like to note that the size of the smallest nanoparticles that we are working with ($\sim$ 15 nm) is of the same order as the particle size where the system may start to show superparamagnetic behavior. (A simple estimate based on the relative strengths of the exchange constant, {\it J} and the bulk value of the  saturation magnetization, {\it $M_S$} for manganites yields the size of a single domain to be $\sim 25-30$ nm). However, in this investigation the specific issue of superparamagnetic  behavior has not been investigated.

\noindent 
\subsection{DEPENDENCE OF TRANSITION TEMPERATURE $T_C$ ON SIZE $d$} 
In Fig. 3 we have shown the non-monotonic variation of $T_C$ with $d$. In the initial stages of size reduction as we reduce the crystal size from the bulk, there is a clear enhancement of $T_C$. However, beyond a certain size ($d < 50$ nm) the $T_C$ rapidly decreases as $d$ is reduced. The data thus clearly show that there are two separate effects operating. One of the effects may dominate over the other in a given size domain. The relative strengths of these two effects will determine the size where the turn around will occur. We propose the following explanation for the observed behavior. 

\noindent 
The first effect that appears to come into play and may dominate in the early stage of size reduction ($d > 100$ nm) is the enhancement of $T_C$ due to the enhancement of the bandwidth ($W$). We find that there is a likely enhancement of $W$ on size reduction. An estimate of the bandwidth $W$ (in the double exchange model) can be made from the from the relation:\cite{bandwidth1,bandwidth2}

\begin{equation}
W \propto cos \omega/d_{Mn-O}^{3.5}
\end{equation} 
where $\omega=\frac{1}{2}(\pi-\theta_{Mn-O-Mn})$. Thus, an increase in the bond angle $\theta_{Mn-O-Mn}$ and a decrease in the bond length $d_{Mn-O}$ would  mean an overall increase in the bandwidth $W$, thereby leading to an enhancement of $T_C$. As has been shown in Fig. 6 and Fig. 7, there are reductions in $\omega$ and $d_{Mn-O}$ on size reduction. Quantitatively, we can estimate the relative change in $W$ using the values of $d_{Mn-O}$ and $\theta_{Mn-O-Mn}$ at $T_C$ for a given size. For $d_{Mn-O}$ and $\theta_{Mn-O-Mn}$ we used the average of the two values for the two oxygen positions. For the smallest particle ($d=15$ nm), the enhancement of $W$ is more than $15\%$ and the particle with $d=50$ nm showing the largest enhancement in $T_C$ ($10\%$) has a relative enhancement in $W$ of $\approx 6\%$. The enhancement of $T_C$ can thus be linked to an enhancement of $W$ on size reduction. A simple phenomenological way to relate the shift in $W$ ($\Delta W$) with $d$ will be to invoke the relation 
$\Delta W \propto d^{-n}$ 
where $n$ is an exponent. One would thus expect that the change in $T_C$ with size reduction arising from the bandwidth effects can be expressed phenomenologically as: 
\begin{equation}
\frac{T_{C}(d)-T_{C}(\infty)}{T_{C}(\infty)} = {\left ( \frac{d_{0}}{d}\right )}^{n}
\end{equation} 
where $T_{C}(\infty)$ and $T_{C}(d)$ are the values of $T_C$ for the bulk sample and for a particle of size $d$. $d_{0}$ is a length scale of nearly atomic dimensions. 

\noindent 
The enhancement of $T_C$ as $d$ is reduced from the bulk (for $d > 50$ nm) observed in $\mathrm{La_{0.67}Ca_{0.33}MnO_3}$ appears to be a special feature of the double exchange ferromagnet that has a large sensitivity to the bandwidth. It is interesting to compare the size dependence of $T_C$ in $\mathrm{La_{0.5}Sr_{0.5}CoO_3}$ which is not a double exchange ferromagnet. We investigated ferromagnetic nanoparticles of $\mathrm{La_{0.5}Sr_{0.5}CoO_3}$ in the size range $d=35-3000$ nm synthesized using a similar chemical route. We found that in this system the behaviour is qualitatively different and there is a monotonic depression of $T_C$ with reduction in $d$. As the particle size was taken down to $35$ nm there was a $20\%$ reduction in $T_C$ from the bulk value. This is unlike the scenario in manganites.  

\noindent 
The second effect that will come into play and will tend to decrease the $T_C$ is the finite size effect that one would expect in magnetic systems of reduced dimensions like in thin films, nanowires and in nanoparticles.\cite{Barber} This occurs when the size becomes comparable to the magnetic correlation length $\xi$ which cannot grow near the critical point due to finite size of the sample. In nanoparticles of elemental ferromagnetic materials this issue has been investigated although for ferromagnetic oxide nanoparticles investigation of size effects on $T_C$ are limited.\cite{Barber} In conventional elemental   ferromagnetic nanowires and nanoparticles this effect has been well established. The value of the correlation length at zero temperature $\xi_{0}$ sets the scale for the finite size effect and for most conventional ferromagnetic systems $\xi_{0} \approx 2-4$ nm.\cite{Sun,Lutz} Generally the finite size effect that leads to depression in the $T_C$ in magnetic systems is the relation:\cite{Barber} 
\begin{equation}
\frac{T_{C}(\infty)-T_{C}(d)}{T_{C}(\infty)} ={\left ( \frac{\xi_{0}}{d}\right )}^{\lambda}
\end{equation} 
As an example of finite size effects in the magnetic transitions in different nano materials, the value of $\lambda$=0.94-0.98 with $\xi_{0}$=2.2-3.35 nm for Ni nanowires.\cite{Sun} Finite size effects have been investigated in oxide films\cite{Andres} of $\mathrm{La_{0.7}Ca_{0.3}MnO_3}$ where the values of $\lambda$ and $\xi_{0}$ are 1.0 and 5 nm respectively. For $\mathrm{La_{0.7}Sr_{0.3}CoO_3}$ films the values are 0.8 and 2.3 nm respectively. 
\noindent
For comparison we investigated the ferromagnetic nanoparticles of $\mathrm{La_{0.5}Sr_{0.5}CoO_3}$ and observed the depression of $T_C$ with $d$ in the size range $d=35-3000$ nm as stated before. From this data we find that $\lambda$ = 0.6 and $\xi_{0}$ = 3.4 nm. This value of $\lambda$ although smaller than other systems, however, is very close to the theoretical predictions of $\lambda\approx 0.7$.\cite{Barber} 

\noindent 
The finite size effects need be separated from the effects that may arise from disordered surfaces which will occur at layers of thickness $\delta$. The finite size effects occur in the bulk or the core of the nanoparticle where the long-range spin order develops and gets truncated by the size of the particle. This needs to be distinguished from effects that arise from disordered spins at the surface. The surface layer with $M_{S}=0$ has $T_{C}=0$ and will not contribute to show any magnetic transition. We did not observe any secondary transition below $T_C$ which will rule out the effect of the surface layer in determination of $T_C$. It acts only to limit the growth of $\xi$ of the spins in the core so that the $T_C$ is depressed by the finite size effect. 

\noindent 
When the above two effects act in tandem, one trying to enhance the Curie temperature (due to enhancement of bandwidth on size reduction) and the other to decrease it (due to finite size effect), one would see the non-monotonic behavior as seen by us. To quantitatively extract the parameters we fit the data shown in Fig. 3 with the phenomenological relation: 
\begin{equation}
\frac{T_{C}(d)-T_{C}(\infty)}{T_{C}(\infty)} ={\left ( \frac{d_{0}}{d}\right )}^{n}-{\left ( \frac{\xi_{0}}{d}\right )}^{\lambda}
\end{equation} 
The fit is shown in the Fig. 3 along with the data. The exponents we obtain for the best fits are $n=1.0$, $\lambda=0.98$ and $\xi_{0}=1.88$ nm and $d_0=1.71$ nm. The fit is reasonable and captures most of the data trend. The maximum deviation of the data from the fit $\delta T_{C}\approx 10$ K occurs near the peak for $d \approx 50$ nm.  The exponents $n$ and $\lambda$ are close to 1, the microscopic spin correlation length $\xi$ is very close to that seen in most ferromagentic materials. The microscopic length $d_0$ is of the order of 2 unit cells. The parameters obtained are all within reasonable physical limits. It thus appears that a combination of bandwidth related enhancement as well as the finite size scaling can describe the dependence of $T_C$ on $d$. The two effects being of similar magnitude, the variation of $T_C$ with $d$ may become very sensitive to factors like preparation condition etc. which can change the balance. 

\subsection{CHANGE IN THE NATURE OF THE PHASE TRANSITION} 
The change in the nature of the ferromagnetic to paramagnetic transition from first order to second order with size reduction is an interesting observation. As stated earlier, the order of the magnetic transition changes from first to second order in size reduced antiferromagnetic MnO. In the case of manganites there are two issues to address which may be interrelated: one is the change in the nature of the transition and the other is the observation that the exponents are close to mean field value.

\noindent 
There can be a number of reasons that can change the nature of the transition. The presence of size distribution can smear the transition region but will not change the nature of the transition. In fact, even in the bulk sample there is a size distribution that is comparable to that in the nanoparticles. But in the bulk sample we see a very clear first order transition. Thus the smearing due to size distribution cannot be the cause of the change of the nature of the transition. This smearing will be like broadening due to temperature fluctuations.\cite{Imry1980}

\noindent
The effect of quenched impurities on first order transition has been studied theoretically\cite{Imry1979} and experimentally\cite{Yusuf} and it is argued that extensive disorder can transform a first order transition into a second order transition and the general criteria has been worked out. While a spread in local transition temperature arising from local disorder/impurity can cause the solid to break into lower temperature phase and a higher temperature phase near the average transition to lower the bulk volume energy, however, this can be prevented by the energy cost of forming interface between such phases which will stabilize the system against local phase fluctuations. In our case, there is no significant disorder or impurity that can cause significant shift in the transition temperature. However, local strain fluctuations can cause fluctuations in the local $T_C$. From the XRD line shape the microstrain, which is the fluctuation in local strain, was found to be $\approx 0.8\%$. Using the bulk modulus ($\sim$ 150 GPa), the local pressure fluctuation can be worked out to be 1.2 GPa. In manganites the transition temperature $T_C$ has a pressure derivative value of $dT_{C}/dP\approx$ 20.2 K/GPa. From this, we can estimate the order of the transition temperature changes arising from the static local strain fluctuations which is $\delta T \approx$ 24 K. This local fluctuation can smear out the transition by producing local phase fluctuations provided the interface energy needed is not large.  The local transition temperature fluctuation that leads to local phase fluctuation occurs over a length scale $l_0$ determined by the interfacial energy, coherence length $\xi_{coh}$ and the latent heat. If $\xi_{coh} > l_0$, the interface energy will stabilize the system against local fluctuations and there will be no significant smearing that can change the nature of the transition. On the other hand, when $\xi_{coh} < l_0$, the transition will be significantly smeared and the nature of the transition can change from first to second order.  However, given the fact that we are dealing with limited system sizes, for small samples $\xi_{coh}$ can be limited by the system size $d$. In that case one may get a situation where $l_0$ can be comparable to $d$ and one can get local phase fluctuations smearing out the transition. In absence of such information like the interfacial energy we cannot reach a definite conclusion whether this local phase fluctuation can turn a first order transition to a second order transition although this a distinct possibility.

\noindent
Another mechanism which is specific to the manganites and can lead to a change in the nature of the transition, is the effect of bandwidth $W$. It has been established that in manganites when the band width is increased without changing the carrier concentration (e.g., by substitution of ions with larger ionic radius in A site,\cite{SrBasubstitution}) the nature of the transition can change from first order to second order. An estimate from the available structural data (for Sr and Ca substituted manganites) can be used to evaluate the change in the bandwidth on Sr substitution in place of Ca, using Eqn. 2. The crossover occurs for rather low Sr substitution ($\sim 10\%$). This corresponds to a change in the bandwidth $W$ $\sim 3-5\%$. In our case, as discussed before, we do find that there is an enhancement in the bandwidth on size reduction which is of similar order. Thus an enhancement of bandwidth which occurs on size reduction can also change the order of the transition. In all probability, the random local strain fluctuations and the bandwidth enhancement act in tandem to change the nature of the transition. 

\noindent
The other issue is that the critical exponents measured from the magnetization data show values that are close to the mean field values. This would be expected in a system of finite size when the Ginzburg coherence length $\xi_{coh}$ cannot grow beyond the system size so that order parameter fluctuations cannot grow so that the system stays close to the mean field region and the critical behavior is not observed.\cite{Chaikin} An estimation of $\xi_{coh}$ can be obtained from the bare microscopic coherence length and the heat capacity change $\Delta C_{v}$ at the transition.\cite{Chaikin} From the depression of $T_C$ with $d$ (Eqn. 5), we obtain the microscopic coherence length $\approx 1.88$ nm. Using experimentally determined values of $C_V$ at the transition, we find that $\xi_{coh}$ can grow to as large as a $\mu$m. Since the particle size is much smaller, the system will stay close to the mean field region.

\section{CONCLUSIONS}
In summary, we have presented an extensive investigation of the effect of size reduction on the ferromagnetic state of $\mathrm{La_{0.67}Ca_{0.33}MnO_3}$ using neutron diffraction along with magnetic measurements. The effects of finite size as seen on some of the physical properties has been analyzed. The analysis of the structural data shows a small but distinct compaction of the $MnO_6$ octahedron with smaller $d_{Mn-O}$ and with $\theta_{Mn-O-Mn}\rightarrow 180^{0}$. This leads to an enhancement of the bandwidth. The neutron data show that even down to a size of $15$ nm the nanoparticles retain ferromagnetic order as can be seen from reduced but finite spontaneous magnetization ($M_S$). The reduction in the spontaneous magnetization on size reduction was explained as arising from a magnetic dead layer with $M_{S}\approx 0$ of thickness $\sim 2$ nm. It is concluded that manganite nanoparticles with size below $5$ nm may not have any $M_S$. The transition to ferromagnetic state changes over from a first order transition to a second order transition with critical exponents approaching mean field values. This was explained as arising from truncation of the coherence length by the finite sample size.  The observed non-monotonic variation of the ferromagnetic transition $T_C$ with size $d$ was explained as arising from simultaneous presence of two effects, one arising from bandwidth enhancement that makes the $T_C$ larger, and the other, a finite size effect that reduces $T_C$. The latter effect wins over at smaller size and the crossover was found to occur in the size range around $50$ nm.

\noindent
\section{ ACKNOWLEDGEMENTS}
The authors would like to thank the Department of Science and Technology, Govt. of India for financial support in the form of a Unit for Nanoscience. One of the authors (TS) also thanks UGC, Govt. of India for a fellowship.


\begin{thebibliography}{100}
\bibitem{TKNath}P.Dey and T.K.Nath, Phys. Rev. B, {\bf73}, 214425 (2006)
\bibitem{NM2}M.M.Savosta, V.N.Krivoruchko, I.A.Danilenko, V.Yu;Tarenkov, T.E.Konstantinova, A.V.Borodin and V.N.Varyukhin, Phys. Rev. B, {\bf69}, 024413 (2004)
\bibitem{NM3}Ning Zhang, Weiping Ding, Wei Zhong, Dingyu Xing AND Youwei Du, Phys. Rev. B, {\bf56}, 8138 (1997)
\bibitem{NM4}Anulekha Dutta, N. Gayathri and R. Ranganathan, Phys. Rev. B, {\bf68}, 054432 (2003)
\bibitem{NM5}S.S.Rao, S.Tripathi, D.Pandey and S.V.Bhat, Phys. Rev. B, {\bf74}, 144416 (2006)
\bibitem{NM6}Anis Biswas and I.Das, Phys. Rev. B, {\bf74}, 172405 (2006)
\bibitem{NM7}T.Zhu, B.G.Shen, J.R.Sun, H.W.Zhao and W.S.Zhan, Appl. Phys. Lett., {\bf 78}, 3863 (2001)
\bibitem{NM8}M.Bibes, Ll.Balcells, J.Fontcuberta, M.Wojcik, S.Nadolski and E.Jedryka, Appl. Phys. Lett., {\bf 82}, 928 (2003)
\bibitem{NM9}J.Curiale, M.Granada, H.E.Troiani, R.D.Sanchez, A.G.Levya, P.Levy and K.Samwer, Appl. Phys. Lett., {\bf 95}, 043106 (2009)
\bibitem{Lopez}M.A.Lopez-Quintela, L.E.Hueso, J.Rivas and F.Rivadulla, Nanotechnology, {\bf 14}, 212 (2003)
\bibitem{MN1}Xavier Batlle and Amilcar Labarta, J. Phys. D: Appl. Phys., {\bf 35}, R15 (2002)
\bibitem{MN2}R.H.Kodama, Journal of Magnetism and Magnetic Materials, {\bf 200}, 359 (1999)
\bibitem{CNRBook}{\it Collosal Magnetoresistance, Charge Ordering and Related Properties of Manganese Oxides} edited by C.N.R.Rao and B. Raveau (World Scientific, 1998)
\bibitem{Shantha}K.Shantha Shankar, Sohini Kar, G.N.Subbanna and A.K.Raychaudhuri, Solid State Comm., {\bf129}, 479 (2004)
\bibitem{Rabinda}R.N.Bhowmik, Asok Poddar, R.Ranganathan and Chandan Majumdar, arXiv:0810.0090v1 [cond-mat.mtrl-sci] 1 Oct 2008
\bibitem{LCMOFilm}M. Ziese, H.C.Semmelhack, K.H.Han, S.P.Sena and H.J.Blythe, J. Appl. Phys., {\bf 91}, 9930 (2002)
\bibitem{PRB}Tapati Sarkar, Barnali Ghosh, A.K.Raychaudhuri and Tapan Chatterji, Phys. Rev. B, {\bf 77}, 235112 (2008)
\bibitem{APL}Tapati Sarkar, A.K.Raychaudhuri and Tapan Chatterji, Appl. Phys. Lett., {\bf 92}, 123104 (2008)
\bibitem{polycrystalline1} J.Mira, J.Rivas, F.Rivadulla, C.Vazquez-Vazquez and M.A.Lopez-Quintela, Phys. Rev. B, {\bf 60}, 2998 (1999)
\bibitem{polycrystalline2}N.Moutis, I.Panagiotopoulos, M.Pissas and D.Niarchos, Phys. Rev. B, {\bf 59}, 1129 (1999)
\bibitem{natureoftransition}C.P.Adams, J.W.Lynn, V.N.Smolyaninova, A.Biswas, R.L.Greene, W.Ratcliff II, S$-$W.Cheong, Y.M.Mukovskii and D.A.Shulyatev, Phys. Rev. B, {\bf 70}, 134414 (2004)
\bibitem{Yusuf}S.Ro$\beta$ler, U.K.Ro$\beta$ler, K.Nenkov, D.Eckert, S.M.Yusuf, K.Dorr and K.-H.Muller, Phys. Rev. B, {\bf 70}, 104417 (2004)
\bibitem{Kim}D.Kim, B.Revaz, B.L.Zink, F.Hellman, J.J.Rhyne and J.F.Mitchell, 
Phys. Rev. Lett., {\bf 89}, 227202 (2002) 
\bibitem{SrBasubstitution}J.Mira, J.Rivas, F.Rivadulla, M.A.Lopez Quintela, Physica B, {\bf 320}, 23 (2002)
\bibitem{XRD}X'Pert PRO Diffractometer, PANanalytical B.V., Lelyweg 1, 7602 EA Almelo, The Netherlands
\bibitem{TEM}JEOL 200 kV, JEOL Ltd., 1 - 2, Musashino 3 - chome Akishima Tokyo 196 - 8558, Japan
\bibitem{SEM}FEI Quanta 200, FEI Company, North America NanoPort 5350 NE Dawson Creek Drive Hillsboro, Oregon 97124 USA
\bibitem{VSM}Lake Shore Cryotronics, Inc., 575 Mc Corkle Blvd, Westerville, OH 43082, USA
\bibitem{WH}G.K.Williamson and W.H.Hall, Acta Metall., {\bf 1}, 22 (1953)
\bibitem{FP}http://www.ill.eu/sites/fullprof 
\bibitem{LSMO}Michael C. Martin, G.Shirane, Y.Endoh, K.Hirota, Y.Moritomo and Y.Tokura, Phys. Rev. B, {\bf 53}, 14285 (1996)
\bibitem{banerjee}B.K.Banerjee, Phys. Lett., {\bf 12}, 16 (1964)
\bibitem{MnO}I.V.Golosovsky, I.Mirebeau, V.P.Sakhnenko, D.A.Kurdyukov and Y.A.Kumzerov, Phys. Rev. B, {\bf 72}, 144409 (2005)
\bibitem{TangJMM}Wei Tang, Wenjian Lu, Xuan Luo, Bosen Wang, Xuebin Zhu, Wenhai Song, Zhaorong Yang, Yuping Sun, J. Magn. Magn. Mater., {\bf 322}, 2360 (2010)
\bibitem{coreshell}P.Kameli, H.Salamati and A.Aezami, J. Appl. Phys., {\bf 100}, 053914 (2006)
\bibitem{Sohini}Sohini Kar, Jayanta Sarkar, Barnali Ghosh and A.K.Raychaudhuri, Phys. Rev. B, {\bf 74}, 085412 (2006)
\bibitem{bandwidth1}M.Medarde, J.Mesot, P.Lacorre, S.Rosenkranz, P.Fischer and K.Gobrecht, Phys. Rev. B, {\bf 52}, 9248 (1995)
\bibitem{bandwidth2}Anthony Arulraj, P.N.Santosh, R.Srinivasa Gopalan, Ayan Guha, A.K.Raychaudhuri, N.Kumar and C.N.R.Rao, J.Phys.:Condens.Matter, {\bf 10}, 8497 (1998)
\bibitem{Barber}M.N.Barber in {\it Phase Transitions and Critical Phenomena} edited by C.Domb and J.L.Lebowitz (Academic, New York, 1983)
\bibitem{Sun}L.Sun, P.C.Searson and C.L.Chien, Phys. Rev. B, {\bf 61}, R6463 (2000)
\bibitem{Lutz}H.Lutz, P.Scoboria, J.E.Crow and T.Mihalisin, Phys. Rev. B, {\bf 18}, 3600 (1978)
\bibitem{Andres}A. de Andres, J.Rubio, G.Castro, S.Taboada, J.L.Martinez and J.M.Colino, Appl. Phys. Lett., {\bf 83}, 713 (2003)
\bibitem{Imry1980}Yoseph Imry, Phys. Rev. B, {\bf 21}, 2042 (1980)
\bibitem{Imry1979}Yoseph Imry and Michael Wortis, Phys. Rev. B, {\bf 19}, 3580 (1979)
\bibitem{Chaikin}{\it Principle of condensed matter physics}, P.M.Chaikin and T.C Lubensky (Cambridge University Press, Cambridge 1995)
\end{thebibliography}
\end{document}